
\input harvmac
\noblackbox\sequentialequations
%
%

\font\ticp=cmcsc10
%
%
\def\dee{\partial}     	\def\lo{\lambda_0}
\def\l{\lambda}		\def\e{\epsilon}
\def\d{\delta}		\def\b{\beta}
\def\k{\kappa}		\def\pt{{\dee\over\dee t}}
\def\ad{a^\dagger}	\def\dl{\langle\kern -.5mm\langle}
\def\bpsi{\bar\psi}     \def\dr{\rangle\kern -.5mm\rangle}
\def\tp{t^{\prime}}	
\def\hpsi{\hat\psi}     \def\a{\alpha}
\def\goto{\rightarrow}  \def\G{\Gamma}
\def\ds{\displaystyle}  \def\ss{\scriptstyle}
%
%
\lref\Doi{M.~Doi, J.~Phys.~A {\bf 9}, 1465 (1976)\semi M. Doi,
J.~Phys.~A {\bf 9}, 1479 (1976).}
\lref\PelitiRev{L.~Peliti, J.~Physique {\bf 46}, 1469 (1985).}
\lref\Peliti{L.~Peliti, J.~Phys.~A {\bf 19}, L365 (1986).}
\lref\Mikhailov{A.~S.~Mikhailov, Phys.~Lett.~A {\bf 85}, 214
(1981)\semi A.~S.~Mikhailov, Phys.~Lett.~A {\bf 85}, 427 (1981).}
\lref\MY{A.~S.~Mikhailov and V.~V.~Yashin, J.~Stat.~Phys. {\bf 38},
347 (1985).}
\lref\Ohtsuki{T.~Ohtsuki, Phys.~Rev.~A {\bf 43}, 6917 (1991).}
\lref\Kang{K.~Kang {\it et al.}, J.~Phys.~A {\bf 17}, L665
(1984).}
\lref\KR{K.~Kang and S.~Redner, Phys.~Rev.~A {\bf 32}, 435 (1985).}
\lref\Kuzovkov{V.~Kuzovkov and E.~Kotomin, Rep.~Prog.~Phys.
{\bf 51}, 1479 (1988).}
\lref\PG{V.~Privman and M.~D.~Grynberg, J.~Phys.~A {\bf 25},
6567 (1992).}
\lref\Lushnikov{A.~A.~Lushnikov, Phys.~Lett.~A {\bf 120}, 135 (1987).}
\lref\Privman{V.~Privman, Phys.~Rev.~A {\bf 46}, R6140 (1992).}
\lref\BL{M.~Bramson and J.~L.~Lebowitz, J.~Stat.~Phys.~{\bf 62}, 297
(1991).}
\lref\Racz{Z.~R\'acz, Phys.~Rev.~Lett. {\bf 55}, 1707 (1985).}
\lref\AF{F.~Family and J.~G.~Amar, J.~Stat.~Phys. {\bf 65}, 1235 (1991).}
\lref\Schulman{L.~S.~Schulman, {\it Techniques and Applications
of Path Integration}, (Wiley, New York, 1981) p.~242 ff.}
\lref\Amit{see for example D. J. Amit, {\it Field Theory, the
Renormalization Group, and Critical Phenomena}, (World Scientific,
Singapore, 1984).}
\lref\CDCi{S.~Cornell, M.~Droz, and B.~Chopard, Phys.~Rev.~A {\bf 44},
4826 (1991).}
\lref\CDCii{S.~Cornell, M.~Droz, and B.~Chopard, Physica A {\bf 188},
322 (1992).}
\lref\CD{S.~Cornell and M.~Droz, Phys.~Rev.~Lett.~{\bf 70}, 3824 (1993).}
\lref\Schnorer{H.~Schn\"orer, I.~M.~Sokolov, and A.~Blumen, Phys.~Rev.~A
{\bf 42}, 7075 (1990).}
\lref\OTB{A.~A.~Ovchinnikov, S.~F.~Timashev, and A.~A.~Belyy, {\it
Kinetics of Diffusion Controlled Chemical Processes}, (Nova Science,
New York, 1989).}
\lref\Privmanii{V.~Privman, (preprint).}
\lref\DS{M.~Droz and L.~Sasv\'ari, Phys.~Rev.~E {\bf 48}, R2343 (1993).}
\lref\FLO{B.~Friedman, G.~Levine, and B.~O'Shaughnessy, Phys.~Rev.~A
{\bf 46}, R7343 (1992).}
\footline={\hfill}
\Title{\vbox{\baselineskip15pt\hbox{}}}
{\vbox{\centerline{Renormalization Group Calculation for the}
\centerline{Reaction $kA\rightarrow\emptyset$}}}
\centerline{\ticp Benjamin P. Lee}
\vskip.1in
\centerline{\sl Department of Physics}
\centerline{\sl Theoretical Physics}
\centerline{\sl 1 Keble Road}
\centerline{\sl Oxford OX1 3NP, UK}
\bigskip\centerline{\bf Abstract}

The diffusion-controlled reaction $kA \goto \emptyset$ is known
to be strongly dependent on fluctuations in dimensions
$d \le d_c=2/(k-1)$.  We develop a field theoretic renormalization
group approach to this system which allows explicit calculation
of the observables as expansions in $\epsilon^{1/(k-1)}$, where
$\epsilon=d_c-d$.  For the density it is found that, asymptotically,
$n \sim A_k t^{-d/2}$.  The decay exponent is exact to all orders
in $\epsilon$, and the amplitude $A_k$ is universal, and is calculated
to second order in $\epsilon^{1/(k-1)}$ for $k=2,3$.  The correlation
function is calculated to first order, along with a long wavelength
expansion for the second order term.  For $d=d_c$ we find
$n \sim A_k (\ln t/t)^{1/(k-1)}$ with an exact expression for $A_k$.
The formalism can be immediately generalized to the reaction
$kA \goto \ell A$, $\ell < k$, with the consequence that the density
exponent is the same, but the amplitude is modified.
\Date{cond-mat/9311064}

\newsec{Introduction}

Diffusion controlled chemical reactions are adequately described by
mean-field type rate equations in higher dimensions, but in lower
dimensions the fluctuations become relevant \refs{\Kuzovkov,\OTB}.
For the reaction $kA\goto\emptyset$ the critical dimension for
fluctuations is conjectured to be
$d_c=2/(k-1)$ \refs{\Kang{,}\KR}.  If $d>d_c$ the density $n(t)$
obeys the rate equation
\eqn\ini{\pt n(t)= -\Gamma n(t)^k,}
with reaction rate constant $\Gamma$.  This implies the density will decay
asymptotically like $n\sim (\Gamma t)^{-1/(k-1)}$.
For $d<d_c$ it is conjectured
on the basis of scaling arguments \refs{\Kang{,}\KR}, rigorous bounds \BL,
and exact results for $d=1$ \refs{\Racz\Lushnikov\AF\Privman{--}\Privmanii},
that $n\sim t^{-d/2}$.  For $d=d_c$ the mean-field power law
with logarithmic corrections is expected.

In this paper we apply renormalization group (RG) methods to this system,
with the goals of verifying the above conjectures and demonstrating
universal quantities.  The formalism developed can be used to calculate
the density and correlation function perturbatively in $\e^{1/(k-1)}$,
where $\e=d_c-d$.  This formalism includes infinite sums for each order
of $\e^{1/(k-1)}$, since the initial density is a relevant parameter and
must be summed to all orders.

Previous work in applying RG to this system was carried out by Peliti
for the case $k=2$ \Peliti.  Using a field theory formulation of this
system, Peliti was able to confirm the conjectured decay exponent, and
also demonstrate that the reactions $A+A\goto\emptyset$ and
$A+A\goto A$ are in the same universality class with regard
to the decay exponent and the upper critical dimension.  Peliti also
made the observations that the coupling constants can be exactly
renormalized to all orders and that there is no wavefunction
renormalization in the theory.  The latter
has the consequence that simple scaling arguments can be used to extract
the decay exponent and the upper critical dimension.  However, these
scaling arguments are not capable of giving other
universal quantities in the system, such as amplitudes or the asymptotic
form of the correlation function.  For these one must do the complete
RG calculation.

Our formalism enables perturbative calculation of these quantities for
general $k$.  For example, we find that the density for $d<d_c$ is given
by $n\sim A_k({\cal D}t)^{-d/2}$ with
\eqn\inii{A_2={1\over 4\pi\e}+{2\ln 8\pi-5\over 16\pi}+O(\e)}
\eqn\iniii{A_3=\left(\sqrt{3}\over 12\pi\e\right)^{1/2}+{9\sqrt{2\pi}
\over 64}+O(e^{1/2}),}
and for $d=d_c$
\eqn\iniv{n(t)\sim\left((k-2)!\over 4\pi k^{1/(k-1)}\right)^{1/(k-1)}
\left(\ln t\over {\cal D}t\right)^{1/(k-1)}.}
where ${\cal D}$ is the usual diffusion constant.

Recent work in applying RG to this system includes that of Ohtsuki
\Ohtsuki, in which the density is calculated, although with
qualitatively different results than those above.  First, Ohtsuki
predicts that the
amplitude for the asymptotic form of the density has the same reaction
rate constant dependence as the mean-field solution:  $n\sim\Gamma^{-1}$
for $k=2$.  Second, the leading order term in the $\e$ expansion for the
density amplitude in \Ohtsuki\ is of order unity.  An RG scheme involving
an external source of particles has been developed by Droz and Sasv\'ari
\DS\ which leads to scaling functions which confirm the decay exponent.
Friedman {\it et al.~}attempted to calculate the density
perturbatively, and concluded that it is necessary to perform a
non-perturbative sum of all orders of $n_0$, the initial density,
when calculating observables \FLO.  This infinite sum is exactly what
we do in our calculation scheme.  To our knowledge there has
been no previous satisfactory, complete RG calculation.

A slightly different field theory formalism for this system was developed
in analogy with bose condensate calculations
\refs{\Mikhailov{,}\MY}.  This approach leads to a confirmation of
the decay exponents as
well.  However, this method is not as readily generalized to an RG
calculation as is the field theory approach of Peliti.

The contents of this paper are as follows.  In section 2 the system
is defined via a master equation.  This is then mapped to a second
quantized representation, and in turn to a field theory.  In section 3
the renormalization of the field theory and the calculation of
observables is addressed.  The latter requires summing infinite sets
of diagrams, for which techniques are developed.
With the formalism established, the density is then calculated in section
4, including correction terms and a discussion of the crossover time scales.
An alternate method for calculating the leading order amplitude, which
does not involve RG, is discussed, and its apparent failure in the case
$k=2$.
In section 5 the correlation function is calculated, and with it universal
numbers for the fluctuations in particle number, both for the total
system and for a small volume $v$.  The local fluctuations in particle
number are
found to be divergent.  Also the second moment of the correlation function
is calculated, giving a correlation length scale.
The case $d=d_c$ is addressed in section 6, and
finally in section 7 a summary these results is given, and the
generalization to $kA\goto\ell A$ is discussed.

\footline={\hss\tenrm\folio\hss}

\newsec{The Model}

Consider a model of particles moving diffusively on a hypercubic lattice
of size $a$,
and having some probability of annihilating whenever $k$
or more particles meet on a lattice site.
This model is defined by a master equation for
$P(\{n\},t)$, the probability of particle configuration
$\{n\}$ occurring at time $t$.  Here \hbox{$\{n\}=(n_1,n_2,\dots,
n_N)$}, where $n_i$ is the occupation number of the $i$th lattice
site.  The appropriate master equation is
\eqn\mei{\eqalign{\pt P(\{n\},t)={{\cal D}\over a^2}
\sum_i\sum_e \biggl\{&(n_e+1)P(
\dots,n_i-1,n_e+1\dots,t)-n_iP(\{n\},t)\biggr\}\cr +\l\sum_i\biggl\{
(n_i&+k)(n_i+k-1)\dots(n_i+1)P(\dots,n_i+k,\dots,t)\cr &-n_i
(n_i-1)\dots(n_i-k+1)P(\{n\},t)\biggr\},}}
where $i$ is summed over lattice sites, and
$e$ is summed over nearest neighbors of $i$.
The first curly brackets piece describes diffusion with diffusion constant
${\cal D}$, and the second annihilation with rate constant $\l$.  The
$P(\{n\},0)$ are given by a Poisson
distribution for random initial conditions with average occupation
number $\bar n_0$.

This master equation can be mapped to a second quantized operator
description, following a general procedure developed by Doi \Doi.
To summarize briefly, operators $a$ and $\ad$ are introduced
at each lattice site, with commutation relations $[a_i,\ad_j]=\delta_{ij}$.
The vacuum ket is given by $a_i|0\rangle=0$.
The state ket of the system at time $t$ is defined to be
\eqn\sqi{|\phi(t)\rangle=\sum_{\{n\}}P(\{n\},t)\prod_i^N(\ad_i)^{n_i}|0
\rangle.}
Then the master equation \mei\ can be written as
\eqn\sqii{-\pt|\phi(t)\rangle=\hat H|\phi(t)\rangle}
with the non-Hermitian time evolution operator
\eqn\sqiii{\hat H=-{{\cal D}\over a^2}\sum_i\sum_e\ad_i(a_e-a_i)
-\l\sum_i\Bigl(1-(\ad_i)^k\Bigr)a_i^k.}
This has the formal solution $|\phi(t)\rangle=\exp(-\hat Ht)|\phi(0)\rangle$.

To compute averages it is necessary to introduce the {\it projection state}
\eqn\sqiv{\langle\>|=\langle 0|\prod_i^Ne^{a_i}.}
Then for some observable $A(\{n\})$,
\eqn\sqv{\dl A(t)\dr\equiv\sum_{\{n\}}A(\{n\})\,P(\{n\},t)=
\langle\>|\hat A\exp(-t\hat H)|\phi(0)\rangle}
where $\hat A$ is the second quantized operator analog of $A$.  Note
that $\langle\>|\ad_i=\langle\>|$.  Therefore any operator $\hat A$
represented in normal ordered form---where all the $\ad_i$ have been
commuted to the left---can be written entirely in terms of the $a_i$.
The operator corresponding to the density is simply $a_i$, while the
correlation function $C(x_i,x_j)$ is given by \hbox{$\ad_ia_i\ad_ja_j$} or
\hbox{$a_i\d_{ij}+a_ia_j$}.  The importance of the $\d$ function term will
be shown later when the renormalized correlation function is calculated.

The second quantized equation can in turn be mapped to
a path integral, with variables $\psi_i$, $\hpsi_i$ at each lattice site,
via the coherent state representation \refs{\PelitiRev,
\Schulman}.
The action corresponding to \sqii\ and \sqiii\ is
\eqn\pii{\eqalign{S[\hpsi,\psi,t]=\sum_i\biggl[&\int_0^tdt\biggl\{\hpsi_i
\dee_t\psi_i-{{\cal D}\over a^2}\hpsi_i
\sum_e(\psi_e-\psi_i)-\l(1-\hpsi_i^k)\psi_i^k
\biggr\}\cr &-\bar n_0\hpsi_i(0)-\psi_i(t)\biggr].\cr}}
The last two terms reflect the Poisson initial conditions and the
projection state.  The path integral form of \sqv\ is then
\eqn\piv{\dl A(t)\dr={\cal N}\int\prod_id\hpsi_id\psi_i A\bigl(\psi(t)\bigr)
e^{-S[\hpsi,\psi,t]}.}
The normalization constant is given by ${\cal N}^{-1}=\int\prod_id
\hpsi_id\psi_ie^{-S[\hpsi,\psi,t]}$.

Next we take the continuum
limit via $\sum_i\goto\int d^dx/a^d$, $\psi_i\goto\psi(x)/a^d$, $\hpsi_i
\goto\hpsi(x)$, $\bar n_0\goto n_0 a^d$, and
$\sum_e(\psi_e-\psi)\goto a^2\nabla^2\psi$.
The initial density is now $n_0$.
The diffusion constant exhibits no singular behavior in the renormalization
of the theory, so it is absorbed into a rescaling of time, giving the action
\eqn\pii{\eqalign{S[\hpsi,\psi,t]=\int d^dx\biggl[&\int_0^tdt\left\{
\hpsi(\dee_t-\nabla^2)\psi-\lo(1-\hpsi^k)\psi^k\right\}\cr
&-n_0\hpsi(0)-\psi(t)\biggr]\cr}}
where $\lo=\l{\cal D}^{-1}a^{(k-1)d}$.

Treating \pii\ as a classical action gives the equations of motion
\eqn\cai{{\delta S\over\delta\hpsi(\tp)}=(\dee_{\tp}-\nabla^2)\psi+k\lo
{\hpsi}^{k-1}\psi^k-n_0\delta(\tp)=0}
and
\eqn\caii{{\delta S\over\delta\psi(\tp)}=-(\dee_{\tp}+\nabla^2)\hpsi+k\lo
({\hpsi}^k-1)\psi^{k-1}-\delta(\tp-t)=0.}
Assuming that $\psi$ and $\hpsi$ are spatially uniform gives the solution
$\hpsi(\tp<t)=1$ and equation \cai\ becomes
\eqn\mfi{\pt\psi=-k\lo\psi^k+n_0\delta(t),}
the mean-field rate equation.  It is consistent that the rate constant
is $k\lo$, since $\lo$ represents the rate at which the reaction occurs,
and the resulting change in particle density is proportional to $k$.
Consider shifting
$\hpsi$ by its classical solution:  $\hpsi\rightarrow 1+\bpsi$.  The action
which results is (up to an overall constant):
\eqn\piii{S[\bpsi,\psi,t]=\int d^dx\left[\int_0^tdt\left\{
\bpsi(\dee_t-\nabla^2)\psi+\sum_{i=1}^k\l_i\bpsi^i\psi^k\right\}
-n_0\bpsi(0)\right],}
where $\l_i={k\choose i}\lo$.  Note that the boundary terms introduced
cancel the $\psi(t)$ term in \pii.

Averages with respect to this action correspond to physical observables,
and are denoted by double brackets.  Single brackets are used for
averages over the curly bracket part
of \piii.  That is, for some observable $A$,
\eqn\piiii{\dl A(x,t)\dr=\left\langle A(x,t)\,e^{n_0\int d^dx\,
\bpsi(x,0)}\right\rangle.}
This is already normalized, since $\langle\exp\{n_0\bpsi(p=0)\}
\rangle=1$.

The dimensions of the various quantities in
\piii, expressed in terms of momentum, are
\eqn\dimi{[t]=p^{-2}\qquad[\bpsi(x)]=p^0\qquad[\psi(x)]=p^d\qquad
[\l_i]=p^{2-(k-1)d}.}
The couplings become dimensionless at the
traditionally accepted value of the critical dimension, $d_c=2/(k-1)$
\refs{\Kang,\KR}.
The relative dimensions of $\psi$ and $\bpsi$ are arbitrary, but this
choice is the most natural.  Any other choice of dimensions would
introduce $n_0$ dependence into the projection state, and cause the
couplings $\l_i$ to have different dimensions.

\newsec{Renormalization of Observables}

The scheme developed for renormalizing the theory follows conventional
RG analysis \Amit.  In this vein a renormalized
coupling is introduced, and shown to have a stable fixed point of order
$\e$.  This is the small parameter of the theory, and not $n_0$,
which implies that the computation of observables
requires summing over an infinite set of diagrams, corresponding to all
powers of $n_0$ in \piiii.  This infinite sum must be grouped into
sets of diagrams whose sums give a particular order of the coupling
constant.  It
will be shown below that this grouping is given by the number of loops.
That is, the infinite set of tree diagrams sum to give the leading order
term in the coupling, the one-loop diagrams the next order term, and so on.
However, before addressing the calculation of observables we turn to the
renormalization of the theory.

\subsec{Renormalization}

To renormalize the theory all that is required is coupling constant
renormalization.  This is because the set of vertices in \piii\ allow
no diagrams which dress the propagator, implying there is no wavefunction
renormalization.  As a consequence the bare propagator is the full
propagator for the theory.

To determine which couplings get renormalized one first
needs to identify the primitively divergent vertex functions.
A general correlation function with $\ell$ $\psi$'s and $m$ $\bpsi$'s
has the dimension
\eqn\pci{[\,\langle\psi(1)\dots\psi(\ell)\bpsi(\ell+1)\dots\bpsi(\ell+m)
\rangle\,]=p^{d\ell}}
where $(1)=(x_1,t_1)$.  The Green's function $G^{(\ell,m)}(p_1,s_1,
\dots,p_{\ell+m},s_{\ell+m})$ is calculated by Fourier and Laplace
transforming the correlation function above, and factoring out overall $p$
and $s$ conserving $\delta$ functions.  The dimensions of this quantity
are
\eqn\pcii{[G^{(\ell,m)}]=p^{d+2-2\ell-(d+2)m}.}
The dimensions of the vertex functions $\Gamma^{(\ell,m)}$ are
given by the Green's functions with the $\ell+m$ external propagators
stripped off.
\eqn\pciii{[\Gamma^{(\ell,m)}]=[G^{(\ell,m)}/(G^{(1,1)})^{\ell+m}]
=p^{2-d(m-1)}.}
The vertex functions with $m\le k$ are those which are primitively divergent
for $d\le d_c$.  Since vertices can only connect
$k$ $\bpsi$'s to some number less than or equal to $k$ $\psi$'s, then
it follows that the primitively divergent diagrams have $m=k$ and $\ell\le k$.

A general $\bpsi^i\psi^k$ vertex is renormalized by the set of diagrams shown
in
\fig\renorm{Sum of all the diagrams which contribute to $\l(p,t_2-t_1)$.
Shown here is the case $k=3$, $i=1$.  These diagrams can be summed
exactly, and are the same for all $i$.}.
In these diagrams the propagator $G_0(p,t)=\langle\psi(p,t)
\bpsi(-p,0)\rangle=e^{-p^2t}$ for $t>0$, $G_0=0$ for $t<0$, and is
represented by a plain line.
Note that this sum is the same for all $i$,  that
is all vertices renormalize identically.  This is a reflection of the fact
that there is a only one coupling in the theory.  These diagrams can be
summed to all orders, as noted in \Peliti.  In $(p,t)$ space the
temporally extended vertex function $\l(p,t_2-t_1)$ is given by
\eqn\lami{\eqalign{\l(p,t_2-t_1)=&\lo\d(t_2-t_1)-\lo^2I(p,t_2-t_1)\cr
&+\lo^3\int_{t_1}^{t_2}d\tp I(p,t_2-\tp)I(p,\tp-t_1)-\dots\cr}}
where $I(p,t)$ is the $k-1$ loop integral
\eqn\lamii{I(p,t)=k!\int\prod_i^{k}\left({d^dp_i\over(2\pi)^d}\right)(2\pi)^d
\d\bigl(p-\sum_i^kp_i\bigr)\exp\bigl(-\sum_i^k p_i^2t\bigr).}
The $\d$ function can be written in integral form, which turns the
integral into a product of $k$ Gaussian integrals.  This gives
\eqn\lamiii{I(p,t)=B_kt^{-(k-1)d/2}e^{-p^2t/k}}
where
\eqn\lamiv{B_k={k!\over k^{d/2}}\left(1\over 4\pi\right)^{(k-1)d/2}.}
Taking the Laplace transform, \hbox{$\l(p,s)=\int_0^{\infty}dt
e^{-st}\l(p,t)$}, makes \lami\ a geometric sum:
\eqn\lamv{\l(p,s)={\lo\over 1+\lo B_k \Gamma(\e/d_c)
(s+p^2/k)^{-\e/d_c}},}
where the $d$ and $k$ have been exchanged for $\e$ and $d_c$.
For a general $\bpsi^i\psi^k$ vertex
the $\lo$ in the numerator is replaced by $\l_i={k\choose i}\lo$, and the
denominator is unchanged.  Therefore the small $s$ and $p$ form of the
vertex function is independent of $\lo$ for all $i$.

The vertex function \lamv\ is used to define a renormalized coupling.
Using the momentum $\k$ as a normalization point, we define the
dimensionless renormalized coupling to be
\hbox{$g_R=\k^{2\e/d_c}\l(s,p)\vert_{s=\k^2,p=0}$}, and the dimensionless
bare coupling $g_0=\k^{2\e/d_c}\lo$.
The $\b$ function is defined by
\eqn\bfi{\b(g_R)\equiv\k{\dee\over\dee\k}g_R=-{2\e\over d_c}g_R+{2\e\over d_c}
B_k\Gamma\biggl({\e\over d_c}\biggr)g_R^2.}
It is exactly quadratic in $g_R$ and has a fixed point $\b(g_R^*)=0$ at
\eqn\bfii{g_R^*=\bigl\{B_k\Gamma(\e/d_c)\bigr\}^{-1}.}
The fixed point is of order $\e$.
{}From the definition of $g_R$, \lamv, and \bfii\ it follows that $g_R^{-1}
=g_0^{-1}+g_R^{*-1}$, or
\eqn\bfiii{g_0={g_R\over 1-g_R/g_R^*}=g_R+{g_R^2\over g_R^*}+\dots}
This will be used to exhange an expansion in $g_0$ calculated in perturbation
theory for an expansion in $g_R$.

\subsec{Calculation Scheme}

Next we develop a Callan-Symanzik equation for the theory.  Given a
correlation function
\eqn\csi{F^{(m)}(t,\lo)\equiv\left\langle\psi(x,t)\left(\int d^dy\,
\bpsi(y,t=0)\right)^m\right\rangle}
The lack of dependence on the normalization scale can be expressed via
\eqn\csii{\left[\k{\dee\over\dee\k}+\b(g_R){\dee\over\dee g_R}\right]
F^{(m)}(t,\k,g_R)=0.}
{}From dimensional analysis $[F^{(m)}]=p^{d-md}$, implying
\eqn\csiii{\left[\k{\dee\over\dee\k}-2t\pt-d+md\right]F^{(m)}=0.}
We are interested in the density $n(t,n_0,g_R,\k)=\sum_m
n_0^mF^{(m)}/m!$.
Substituting \csiii\ into \csii, and summing to get the density gives
the equation
\eqn\csiv{\left[2t\pt-dn_0{\dee\over\dee n_0}+\b(g_R){\dee\over\dee g_R}
+d\right]n(t,n_0,g_R,\k)=0.}
This is solved by the method of characteristics, and has the solution
\eqn\csv{n(t,n_0,g_R,\k)=(\k^2t)^{-d/2}n\Bigl(\k^{-2},\tilde n_0(\k^{-2}),
\tilde g_R(\k^{-2}),\k\Bigr),}
with the characteristic equations for the running coupling and initial
density
\eqn\csvi{2t{\dee\tilde n_0\over\dee t}=-d\tilde n_0\qquad\tilde n_0(t)=n_0,}
\eqn\csvii{2t{\dee\tilde g_R\over\dee t}=\b(\tilde g_R)\qquad
\tilde g_R(t)=g_R.}
Because of the simple form of the $\b$ function, the running coupling can
be found exactly:
\eqn\csviii{\tilde n_0(\tp)=(t/\tp)^{d/2}n_0,}
\eqn\csix{\tilde g_R(\tp)=g_R^*\left(1+{g_R^*-g_R\over g_R
(t/\tp)^{\e/d_c}}\right)^{-1}.}
One then sets $\tp=\k^{-2}$ and plugs the result into \csvi.  Notice
that in the large $t$ limit $\tilde g_R\rightarrow g_R^*$.

In conventional RG analysis the mechanics developed above is used in the
following way:  one calculates an expansion in powers of
$g_0$, and then converts this to an expansion in powers of $g_R$ via
\bfiii.  As long as the expansion coefficients are non-singular in $\e$,
then the $g_R$ expansion can be related to an $\e$ expansion via \csv.
That is, we substitute $t\goto\k^{-2}$, $n_0\goto\tilde n_0$,
$g_R\goto\tilde g_R$, in the $g_R$ expansion, and multiply
by the overall factor shown in \csv.  Then for large $t$,
\hbox{$\tilde g_R\goto g_R^*$}
giving $n(t,n_0,\lo)$ as an expansion in powers of $\e$.  For a given
coefficient in the
$g_R$ expansion we keep only the leading term for large $n_0$, since
$\tilde n_0\sim t^{d/2}$ and so the subleading terms in $\tilde n_0$ will
correspond to sub-leading terms in $t$.

The identification of the leading terms in $g_0$ is less straightforward
than it is in conventional RG calculations, since  the sum over all
powers of $n_0$ must be taken into account.  For the density,
tree diagrams are of order $g_0^in_0^{1+i(k-1)}$ for integer $i$.
Diagrams with $j$ loops are of order $g_0^in_0^{1+i(k-1)-j}$.  Since
the addition of loops makes the power of $g_0$ higher relative to the
power of $n_0$, we hypothesize that the number of loops will serve as
an indicator of the order of $g_0$.  This will be shown to be the case
via explicit calculation.

\subsec{Tree Diagrams}

To calculate all possible diagrams of a given number of loops it is
necessary to develop two tree-level quantities:  the
classical density and the classical response function.
The term classical means averaged with respect to the classical action,
which is the action \piii, but with only the $\bpsi\psi^k$ vertex.
The classical
density is given by sum of all tree diagrams which terminate with a
single propagator, as shown in
\fig\clden{The classical density, represented as a dashed line, is given
by (a) the complete sum of tree diagrams, and (b) an integral equation.
The latter is equivalent to the mean-field rate equation.  Shown here
is the case $k=2$.}, and is represented graphically by a dashed line.
These diagrams are evaluated in momentum space.  From \piiii\ it
follows that the $\bpsi(t=0)$ in the initial state all have $p=0$,
so all diagrams at tree level have $p=0$.

Shown also in \clden\ is an exact graphical relation for the infinite
sum, which is equivalent to the mean-field rate equation \mfi.  This
can be seen by considering the diagram in position space, and
 acting with $(\dee_t-\nabla^2)$, the inverse of the
Green's function $G_0$, on either side of the diagramatic equation.
Note that the combinatoric factors involved in attaching the full density
lines to vertices is different than for propagators, which is discussed
in appendix A.
This equation has the exact and asymptotic (large $t$) solutions
\eqn\mfii{n_{cl}(t)={n_0\over(1+k(k-1)n_0^{k-1}\lo t)^{1/(k-1)}}\sim
\left(1\over k(k-1)\lo\right)^{1/(k-1)}t^{-1/(k-1)}.}
The asymptotic solution depends on the coupling strength, but not the
initial density.

The response function is defined by
\eqn\rfi{G(p,t_2,t_1)\equiv\dl\psi(-p,t_2)\bpsi(p,t_1)\dr,}
and the
classical response function is the above quantity with only tree diagrams
included in the averaging.
It is represented graphically by a heavy line,
and is given by the sum of diagrams as shown in
\fig\response{The response function, shown as a heavy line, is given as a
sum of the bare propagator plus a term with a single vertex connecting
$k-1$ full density lines, plus a term with two vertices, and so on.  Shown
here is $k=3$.  These diagrams can be summed exactly.}.
Note that the only $p$-dependence is that of the bare propagator. That is,
the density lines all carry no momentum.  The
time dependence of the propagators connecting the vertices cancels to leave
only overall dependence on $t_1,t_2$.  The vertices are now symmetric
under interchange, so we can trade the requirement that they be ordered
for a factor of $1/n_v!$ where $n_v$ is the number of vertices.  The sum
of diagrams is then identified as the Taylor expansion of an exponential,
giving
\eqn\rfii{\eqalign{G_{cl}(p,t_2,t_1)=&e^{-p^2(t_2-t_1)}\exp\left\{-k^2\lo
\int_{t_1}^{t_2}dt\>n_{cl}(t)^{k-1}\right\}\cr =&e^{-p^2(t_2-t_1)}
\left(1+k(k-1)n_0^{k-1}\lo t_1\over 1+k(k-1)n_0^{k-1}\lo t_2\right)^{k/(k-1)}.
\cr}}
The extra factor of $k$ associated with each $-k\lo$ vertex
is a consequence of the combinatorics (see Appendix A).
{}From \piiii\ it follows that $\dl\psi(t)\bpsi(0)\dr=
\dee\dl\psi(t)\dr/\dee n_0$
or $G(p=0,t,0)=\dee n(t)/\dee n_0$.  This relation should also hold for the
classical density and response function, as is the case for the solution above.

\newsec{Density Calculation}

With the classical or tree-level solutions of the previous section, and
the renormalization scheme developed above,  the
asymptotic form of the density can now be calculated.  The solution for
the tree diagrams in terms of $g_0$, or $\lo$, is given by \mfii.  To
leading order in $g_R$ one just replaces $\lo$ with $g_R\k^{2\e/d_c}$.
For large $t$ the running coupling $\tilde g_R\goto g_R^*$, which gives
\eqn\deni{n^{(0)}(t)={n_0\over (1+k(k-1)n_0^{k-1}
g_R^*\,t^{(k-1)d/2})^{1/(k-1)}}.}
The superscript on the density refers to the number of loops in the
calculation.  The asymptotic form of this expression is
\eqn\denii{n^{(0)}(t)\sim\left((k-2)!\over 2\pi(k-1)k^{1/(k-1)}
\e\right)^{1/(k-1)}t^{-d/2}+O(\e^{1-1/(k-1)}).}
The term in parentheses is the leading order term in $A_k$, the amplitude
of the $t^{-d/2}$ component of the density.

\subsec{Amplitude Corrections for $k=2$}

Next the corrections from the higher loop diagrams are calculated.
It will be shown that adding a loop makes the sum of diagrams
an order $g_R^{1/(k-1)}$ higher.  At $k-1$ loops the diagrams will contain
a singularity in $\e$, caused by the appearance of the first primitively
divergent diagram.  However this singularity is cancelled when the $g_R^2$
correction to $g_0$ in \bfiii\ is included in the tree diagram sum.
In general the higher order terms in \bfiii\ will cancel all divergences
in the coefficients of the $g_R$ expansion.  This will be illustrated
in the one-loop corrections for $k=2$.

The infinite sum of all one-loop diagrams can be written in terms of the
classical response function found above.  The sum of diagrams is shown in
\fig\two{One- and two-loop diagrams for $k=2$. By using
the response function all such diagrams are included.  Diagram (a) is
used to calculate the amplitude correction.}.
Expressing this graph in integral form
\eqn\aci{\eqalign{n^{(1)}(t,&n_0,g_0,\k)=\cr
&2\int dt_2dt_1{d^dp\over (2\pi)^d}G_{cl}(0,t,t_2)
(-2\lo )G_{cl}(p,t_2,t_1)^2(-\lo )n_{cl}(t_1)^2,\cr}}
where the time integrals are over $0<t_1<t_2<t$.
Taking the large $n_0$ limit of \aci\ to extract the asymptotic part gives
\eqn\acii{n^{(1)}(t,n_0,g_0,\k)={1\over t^2}\int dt_2dt_1{d^dp\over
(2\pi)^d} t_2^{-2}e^{-2p^2(t_2-t_1)}t_1^2+O(n_0^{-1}).}
Notice that this is independent of $g_0$, consistent with the prediction that
the one-loop diagrams are of order $g_R^0$ and provide a correction to
the leading term in \denii.
The integral can be done exactly.  Expressing the leading piece in terms of
$g_R^*$, and the rest as an expansion in $\e$:
\eqn\aciii{n^{(1)}(t,n_0,g_0,\k)=t^{-d/2}\left({1\over 2g_R^*}-{2C+5\over
16\pi}+O(\e)\right),}
where $C$ is Euler's constant.
The correction to the tree-level component due to the subleading term in
$g_0(g_R)$ is
\eqn\aciv{n^{(0)}(t,g_R)={1\over 2g_R}t^{-d/2}-{1\over 2 g_R^*}t^{-d/2}
+O(g_R).}
The singular parts of the $g_R^0$ coefficient cancel as advertised.
Combining \aciii\ and \aciv\
and making use of the Callan-Symanzik solution \csv\ gives
\eqn\acv{A_2={1\over 4\pi\e}+{2\ln 8\pi-5\over 16\pi}+O(\e).}

The two-loop diagrams are also shown in \two.
They all contribute to order $g_R^1$.  Unfortunately we are unable
to evaluate diagrams (f,g) due to the complicated time dependence of
the vertices, which prohibits calculation of the $O(\e)$ term in
$A_2$.  The most singular of the diagrams, (b-d), are of order $\e^{-2}$.
These diagrams can be calculated and the singular pieces cancel as
expected.

Note that the asymptotic, or large $n_0$, limits of the classical density
and the classical response function are of order $n_0^0$, which implies
that the asymptotic time dependence of the density, calculated to any number of
loops, will be $t^{-d/2}$.  Therefore the decay exponent is exact to all
orders in $\e$.

The cancellation of the singularities which appear in the $g_R$ expansion
can be most easily understood by viewing the correction terms in \bfiii\
as counterterms introduced to cancel primitive divergences.  That is,
considering $\d g_R=g_R^2/g_R^*+O(g_R^3)$, and calculating the first
order term in $\d g_R$ at tree level gives a diagram similar to \two\ (a),
but with the counterterm in place of the loop.  This diagram, when added
to the one-loop diagram, cancels the singularity in the $g_R^0$ coefficient.
Two-loop diagrams (b-f) can be viewed as primitively divergent loops
added to the one-loop diagram (a).  The order $\d g_R$ terms in
the one-loop diagram are equivalent to diagrams (b-f) with a counterterm
in place of the additional loop, and will cancel the divergences in these
diagrams.  Diagram (g) differs in that it is not a primitively
divergent loop `added on'  to diagram (a), but it is also non-singular.

\subsec{Amplitude Corrections for $k=3$}

The one- and two-loop diagrams for $k=3$ are shown in
\fig\three{One- and two-loop diagrams  for $k=3$.  Diagram (a) contains
no $\e$ singularity, and is used to calculate the amplitude correction.}.
The one-loop diagram contains no singularity, and gives the order $g_R^0$
correction to \denii.  The asymptotic piece is given by the integral
\eqn\acvi{n^{(1)}(t,n_0,g_0,\k)={3\over 2t^{3/2}}\int dt_2dt_1{d^dp\over
(2\pi)^d} t_2^{-2}e^{-2p^2(t_2-t_1)}t_1^{3/2}+O(n_0^{-1}).}
Performing the integral and using \csv\ we find the amplitude
\eqn\acvii{A_3=\left(\sqrt{3}\over 12\pi\e\right)^{1/2}+{9\sqrt{2\pi}
\over 64}+O(e^{1/2}).}

The two-loop diagrams are of order $g_R^{1/2}$,
although, similar to the case of $k=2$, we are unable to calculate
diagrams  (f-i).  The only diagram with a singularity is (j) which can
be calculated to demonstrate
that the $g_R^{1/2}$ coefficient is non-singular as expected.

\subsec{Dressed Tree Calculation}

There exists an alternate method for calculating the leading order
amplitude of the density which does not require using the RG formalism.
However, there is a discrepancy between this method, the dressed tree
sum, and the RG in the case $k=2$.  We present the dressed tree
calculation below, and an
explanation for why we believe the RG to be correct for $k=2$.

Consider summing the most divergent diagrams for each power of $\lo$ and
$n_0$.  This is equivalent to summing the dressed tree diagrams, which
are tree diagrams with all the vertices replaced by
the temporally extended vertex function \lami.
The sum of these diagrams, $n_{dt}(t)$, satisfies the diagramatic
equation shown in
\fig\dressedtree{Exact diagramatic equation for $n_{dt}(t)$, the sum of the
dressed tree diagrams.},
where $n_{dt}$ is represented by a dotted line.
As with the tree diagram sum, acting on this equation with the propagator
inverse $(\dee_t-\nabla^2)$ gives a differential equation
\eqn\dti{\dee_tn_{dt}(t)=n_0\d(t)-k\int_0^td\tp\l(p=0,t-\tp)n_{dt}(\tp)^k.}
Laplace transforming the equation gives
\eqn\dtii{sn(s)-n_0=-k\l(0,s)n^k(s),}
where $n(s)=\int_0^{\infty}dte^{-st}n(t)$ and $n^k(s)=\int_0^{\infty}dt
e^{-st}n(t)^k$.  The transform of the vertex function $\l(0,s)$ is known
exactly, and
is given by \lamv.  However, the equation is not algebraic in
$n(s)$, making it difficult to obtain an exact solution.  To proceed,
we assume $n_{dt}\sim\tilde At^{-\a}$, so that for small $s$,
$n(s)\sim\hbox{$\tilde A\Gamma(1-\a)s^{\a-1}$}$.  Also, $n^k(s)\sim\hbox{
$\tilde A^k\Gamma(1-k\a)s^{k\a-1}$}$, and $\l(0,s)\sim\hbox{$s^{\e/d_c}/
(B_k\Gamma(\e/d_c))$}$.
The transform of $n_{dt}(t)^k$ is calculated by imposing a small $t$
regulator, which is justified as the transform of the exact solution
does exist, and then taking the small $s$ limit.  The amplitude which
results is independent of the regulator.
Substituting these in to \dtii\ and taking the small $s$ limit of the
eqation gives $\a=d/2$, and the amplitude
\eqn\dtiii{\tilde A^{k-1}={B_k\G\left(\ds 2\e\over\ds k-1\right)\G\left({\ds
k-2\over\ds k-1}+{\ds\e\over\ds 2}\right)\left({\ds 1\over\ds k-1}-{\ds
k\e\over\ds 2}\right)\over
k\,\G\left({\ds k-2\over\ds k-1}+{\ds k\e\over\ds 2}\right)}.}
For $k\ne 2$ the non-singular $\G$ functions cancel to leading order in
$\e$, with the result $\tilde A=A_k+O(\e^0)$.  However, for $k=2$ all
the $\G$ functions are singular, which has the consequence that
$\tilde A_2=2A_2+O(\e^0)$.  In light of this, it seems necessary to
find an explanation why this particular set of diagrams sums to give
the proper leading order term for general $k$, but not for $k=2$, if
indeed the RG is giving the correct leading order term.

Consider the set of dressed one-loop diagrams.  That is, the set of
diagrams given in \two\ (a) and \three\ (a), but again with each
vertex replaced by the temporally extended vertex function.  While
it would be difficult to calculate this sum, it is possible to see
a property specific to $k=2$ that they have.  The analog of the
classical densities in these diagrams is the dressed tree
density $n_{dt}\propto t^{-d/2}$.  Therefore for general $k$ there
is a time integral over $t^{-kd/2}$, or $t^{-k/(k-1)-k\e/2}$.  This
time integral will be in the form of a Laplace convolution integral,
similar to \dti.  Using a regulated transform as before, the amplitude
of the small $s$ limit will be proportional to \hbox{$\G\bigl((k-2)
/(k-1)+k\e/2\bigr)$}.
For $k\ne 2$ this is non-singular at $\e=0$, but for $k=2$ it is of
order $\e^{-1}$.  Therefore these diagrams
are part of the leading order amplitude for $k=2$.  As a result, it
would appear that the discrepancy is a consequence of the failure of
the dressed tree method, and not of the RG.

\subsec{Crossovers}

There are two crossover time scales in this system, one given by
$n_0$ and one by $\lo$.  For the coupling
constant crossover we consider the large $t$ expansion of \csix\
\eqn\cci{\tilde g_R=g_R^*\left(1-\lo^{-1}t^{-\e/d_c}
+O(t^{-2\e/d_c})\right).}
Including the correction term in the density calculation will
generate a $\lo$ dependent term proportional to
$t^{-d/2-\e/d_c}$.  From \cci\ it follows that the characteristic crossover
time is given by $t_{\lo}\sim(\e/\lo)^{d_c/\e}$.  In terms of
the constants in the master equation, \hbox{$t_{\l}\sim a^2{\cal D}^{-1}
(\e {\cal D}/a^2\l)^{d_c/\e}$}.  For small $\e$, or
large $\lo$, the time required to
reach the fluctuation dominated regime becomes small.

The $n_0$ crossover is calculated by keeping the order $n_0^{-1}$ terms
in the integrals performed above.  These terms will pick up an extra
factor of $t^{-d/2}$ when put into \csv, so
the exponent of the leading $n_0$ dependent term in the density is $t^{-d}$.
The characteristic crossover time is only weakly $\e$ dependent, and is
given by $t_{n_0}\sim {\cal D}^{-1}n_0^{-2/d}=a^2{\cal D}^{-1}\bar n_0^{-2/d}$.

If the $n_0$ crossover occurs first, then
for intermediate times $t_{n_0}\ll t\ll t_{\lo}$ one would expect the system
to obey the asymptotic form of the mean-field solution.  That is,
$n\sim[k(k-1)\lo t]^{-1/(k-1)}$.  If the $\lo$ crossover occurs first
it is less clear what the behavior in the intermediate regime will be.
The contribution from the tree diagrams will be exactly \deni, which
does not become a power law until the $n_0$ crossover is reached.  This
is complicated even further by the higher order diagrams.

\newsec{Correlation Function Calculation}

The density correlation function is given by
\eqn\aai{C(x,t)=\dl\bigl(\psi(x,t)+\d^d(x)\bigr)\psi(0,t)\dr.}
where the $\d$ function is a consequence of the second quantized
operators developed in section 2.
A Callan-Symanzik equation for the correlation function can be developed
in a similar fashion as before.  Consider the function
\eqn\cfi{F^{(m)}(p,t,\lo)\equiv\int d^dx\,e^{-ip\cdot x}\left\langle
\bigl(\psi(x,t)+\d^d(x)\bigr)\psi(0,t)\left(\int d^dy\,\bpsi(y,t=0)
\right)^m\right\rangle.}
Dimensional analysis gives $[F^{(m)}]=p^{d-md}$.
The correlation function $C(p,t)$ is given by $\sum_m n_0^mF^{(m)}/m!$.
This leads to the equation
\eqn\cfii{\left[2t\pt-p{\dee\over\dee p}-dn_0{\dee\over\dee n_0}+
\b(g_R){\dee\over\dee g_R}+d\right]C(p,t,n_0,g_R,\k)=0,}
which has the solution
\eqn\cfiii{C(p,t,n_0,g_R,\k)=(\k^2t)^{-d/2}C\Bigl(\tilde p(\k^{-2}),t=\k^{-2},
\tilde n_0(\k^{-2}),\tilde g_R(\k^{-2}),\k\Bigr),}
with $\tilde g_R$ and $\tilde n_0$ given by \csviii\ and \csix, and
\eqn\cfiv{\tilde p(\tp)=p\sqrt{t\over\tp}.}

Again the calculation of the right hand side of \cfiii\ is divided into
the number of loops.  First the connected and disconnected pieces are
separated
\eqn\cfv{C(p,t)=n(t)+g(p,t)+\d^d(p)n(t)^2.}
The first term on the right hand side is a consequence of the $\d$ function
in \aai, and is considered part of the connected correlation function.
The disconnected tree-level graphs are of the order $g_0^in_0^{2+i(k-1)}$,
and represent the leading order terms in the correlation function.  This
is reasonable, as the classical solution of this system corresponds to
the absence of correlations.  The connected tree-level diagrams, which
are the leading terms in $g(p,t)$, are of
order $g_0^in_0^{1+i(k-1)}$, and represent the leading corrections due
to fluctuations.  The tree-level and one-loop diagrams for $g(p,t)$ in
the case $k=2$ are shown in
\fig\corrtwo{The diagrams for the connected correlation function at tree
level and one loop, for $k=2$.}.
Diagram (a) can be calculated explicitly to give the leading term
\eqn\cfvi{\eqalign{g(p,t)&=-{1\over 4\pi\e}t^{-d/2}f_2(p^2t)+O(\e^0)\cr
f_2(x)&=-{e^{-2x}\over 4x^3}+{1\over 4x^3}-{1\over 2x^2}+
{1\over 2x}.\cr}}
The function $f_2(x)$ is regular at $x=0$, with $f_2(0)=1/3$.  For large
$x$, $f_2(x)\sim 1/(2x)$.

We are unable to evaluate the one-loop diagrams analytically for general
$p$, although it is possible to calculate an expansion in $p^2$, which
we have done to order $p^2$.
For the connected correlation function, $\bar C(p,t)=n(t)+g(p,t)$,
\eqn\aaii{\bar C(p,t)=\left[{1\over 6\pi\e}+{9\ln 8\pi-26\over 108\pi}
+\left({1\over 24\pi\e}+{15\ln 8\pi-19\over 720\pi}\right)p^2t+\dots
\right]t^{-d/2}+O(\e).}
With the expansion above it is possible to calculate the second moment
of $\bar C(x,t)$, giving a length scale for the correlations.
For $\bar C(p,t)=A+Bp^2+\dots$ the second moment
$-\xi^2\equiv\int d^dx\,x^2\bar C(x,t)/\int d^dx\bar C(x,t)=-2B/A$.
The negative sign in the definition
of $\xi$ is required since the second moment is negative, indicating
that the particles are negatively correlated at larger distances.
For $k=2$ the length $\xi$ is given by
\eqn\xii{\xi_2=\sqrt t\left({\sqrt 2\over 2}+{73\sqrt 2\over 360}\e+O(\e^2)
\right).}

The correlation function can be used to calculate the fluctuations in
the density.  For example, the fluctuations in the local density are
given by integrating $C(p,t)$ over $p$.  However, the $p$-independent
term causes this integral to diverge.  One can consider the fluctuations
of the average particle number of fiducial volume $v$.  This is given by
\eqn\fvi{(\d N_v)^2=v\int_vdx\,\bar C(x,t)=vn(t)+O(v^2),}
where translational invariance is assumed.  The order $v$ contribution
originates from the $\d$ function in \aai.  For small $v$ the
fluctuations go as $\d N_v\sim\sqrt{vn(t)}$, which is universal.
Also, $\d N_v/N_v\sim 1/\sqrt{vn(t)}$, which diverges as $v$ goes to
zero, consistent with the local fluctuations being divergent.

The fluctuations in the total number of particles is given by
$V\bar C(p=0,t)$ where $V$ is the volume of the system.
When divided by the square of the average number of particles,
$V^2n(t)^2$, this gives
\eqn\aaiii{{(\d N)^2\over N^2}V=\left({8\pi\over 3}\e-{36\pi\ln 8\pi-
76\pi\over 6}\e^2+O(\e^3)\right)t^{-d/2}.}
Note that all these fluctuation terms would be negative if the $\d$
function term were neglected.  That is, $\dl\psi(x)^2\dr<0$, a
demonstration that the fields introduced via the path integral
formulation of \PelitiRev\ are complex.

The diagrams contributing to $g(p,t)$ for $k=3$ are shown in
\fig\corrthree{The diagrams for the connected correlation function at
tree level and one loop, for $k=3$.}.
The leading order term for the connected part is
\eqn\cfviii{\eqalign{g(p,t)&=-\left(\sqrt 3\over 12\pi\e\right)^{1/2}t^{-d/2}
f_3(x)+O(\e^0)\cr f_3(x)&={e^{-2x}3\sqrt{\pi}{\rm erfi}(\sqrt{2x})\over
16\sqrt{2} x^{5/2}}-{3\over 8x^2}+{1\over 2x}\cr}}
where ${\rm erfi}(x)=-i\,{\rm erf}(ix)=(2/\sqrt\pi)\int_0^xdye^{y^2}$.
The function $f_3(x)$ is also regular,
with $f_3(0)=2/5$ and $f_3(x)\sim 1/(2x)$ for large $x$.

The one-loop diagrams can be calculated as an expansion in $p^2$, with
the net result
\eqn\aaiv{\eqalign{\bar C(p,t)=&\Biggl[{3\over 10}\left(\sqrt 3\over 3
\pi\e\right)^{1/2}+{81\sqrt{2\pi}\over 1600}+{288\over 875}\sqrt{2\over\pi}
\cr &+\left({4\over 35}\left(\sqrt 3\over 3\pi\e\right)^{1/2}
+{9\sqrt{2\pi}\over 98}-{27546\over 42875}\sqrt{2\over\pi}\right)
p^2t+\dots\Biggr]t^{-d/2}+O(\e)\cr}}
In this case the sign of the second moment of the correlation function
depends on $\e$.  For $\e<0.4$ the second moment is negative, and the
resulting length scale is given by
\eqn\xiii{\xi_3=\sqrt{t}
\left({4\over\sqrt{21}}+\left(2\sqrt 3\over 21\right)^{1/2}
\left({711\pi\over 2240}-{2217\over 490}\right)\e^{1/2}+O(\e)\right).}
The fluctuations in total particle number are given by
\eqn\aav{{(\d N)^2\over N^2}V=\left({6\sqrt\pi\over 5}\e^{1/2}+
(2\pi\sqrt 3)^{1/2}\left({1152\over 875}-{189\pi\over 400}\right)\e
+O(\e^{3/2})\right)t^{-d/2}.}

\newsec{$d=d_c$}

In general, when $d<d_c$, certain relevant parameters determine the
critical exponents of the system.  When $d=d_c$ these parameters
become marginally irrelevant.  In such a case the exponents are
given by mean-field theory, but with logarithmic corrections.  In our system
the marginally irrelevant parameter is the coupling $\lo$.

When $d=d_c$ the Callan-Symanzik solution \csv\ still holds, although with
a different running coupling.  The $\b$ function can be calculated either
with a cutoff which is taken to infinity or by taking $\e\goto 0$ in \bfi\
with the same result: \hbox{$\b(g_R)=2B_kg_R^2$}.  This gives the running
coupling
\eqn\dci{\tilde g_R(\k^{-2})={g_R\over 1+g_RB_k\ln(\k^2t)}.}
For large $t$ the coupling goes to zero, which is the only fixed point
of the $\b$ function.  Using the asymptotic form $\tilde g_R\sim
\{B_k\ln(\k^2t)\}^{-1}$ in the tree-level sum gives
\eqn\dcii{n(t)\sim\left((k-2)!\over 4\pi k^{1/(k-1)}\right)^{1/(k-1)}
\left(\ln t\over t\right)^{1/(k-1)}\left[1+O\bigl((\ln t)^{-1/(k-1)}
\bigr)\right].}
Higher order terms in $\tilde g_R$ will give sub-leading time dependence,
so this represents the full leading order amplitude.  Notice that the
correction terms are only an order $(\ln t)^{-1/(k-1)}$ smaller, which
will make
time required to reach the asymptotic regime large.

The same procedure gives an exact expression for the leading term in
the correlation function as well.  For $k=2$
\eqn\dciii{\bar C(p,t)={1\over 8\pi}\Bigl(1-f_2(p^2t)\Bigr)\left(\ln t
\over t\right)\left[1+O\bigl((\ln t)^{-1}\bigr)\right]}
and for $k=3$
\eqn\dciv{\bar C(p,t)=\left(\sqrt 3\over 12\pi\right)^{1/2}
\Bigl(1-f_3(p^2t)\Bigr)\left(\ln t
\over t\right)^{1/2}\left[1+O\bigl((\ln t)^{-1/2}\bigr)\right].}

\newsec{Summary and Generalization to $kA\goto\ell A$}

With the RG calculation developed above we are able to calculate
various universal quantities for this system.  These include the
amplitude of the asymptotic density for $d\le d_c$, given by
\acv, \acvii, and \dcii, and the connected correlation function.
Also universal are the fluctuations in total particle number and
the fluctuations in particle number in a small volume $v$.

The density amplitude for $k=2$ can be compared to the exact solution
for $d=1$ of $A_2=(8\pi)^{-1/2}\approx 0.20$ \Lushnikov.  Putting $\e=1$
in our expansion yields $A_2=0.08 + 0.03+\dots$.  The agreement is less
than satisfactory, indicating that the $\e$ expansion will
not be quantitatively accurate to $\e=1$.  However, the $\e$ expansion
provides the only systematic derivation of universality and scaling.

Our results can be immediately generalized to a coagulation
reaction $kA\goto\ell A$, $\ell<k$.  The only change in the field
theory is the vertices $\l_i$ in \piii:
\eqn\disci{\l_i=\cases{\lo{k\choose i}-\lo{\ell\choose i},&$i\le\ell$\cr
&\cr\lo{k\choose i}, &$i>\ell$.\cr}}
The renormalization follows identically.  For example, the leading term
in the amplitude, given by \denii, is generalized to
\eqn\discii{A_{k,\ell}=\left(k\over k-\ell\right)^{1/(k-1)}A_k+O(\e^0).}
This proportionality is not generally true for all terms in the $\e$
expansion, although it does happen to hold when $k=2$.  To see this
consider a rescaling $\psi\goto b\psi$, $\bpsi\goto\bpsi/b$, and
$n_0\goto bn_0$ in the action \piii.  The only terms changed by such a
rescaling are the couplings $\l_i\goto b^{i-k}\l_i$, which for $k=2$ is
only the coupling $\l_1$.  Starting from the theory $A+A\goto\emptyset$
and making the scale transformation with $b=2$ gives exactly the theory
for $A+A\goto A$.  As a
consequence, the density for $A+A\goto A$, starting from an
intial density of $n_0$, will for all times be exactly twice the density
of the system $A+A\goto\emptyset$ with initial density of $n_0/2$.
This result agrees with the recent exact solution of a particular
model of $A+A\goto (\emptyset,A)$ in $d=1$ \Privmanii, although it
should be noted that this relation is not truly universal for all times,
 as it only holds when the irrelevant couplings are excluded.
The asymptotic amplitude is universal, and so the relation
$A_{2,1}=2A_{2,0}$ is exact to all orders in $\e$, and independent
of the initial densities.

For $k=3$ such a simple relation does not hold.  We can consider all
three theories, $\ell=0,1,2$, combined with relative strengths
$r_0, r_1, r_2$, where $\sum_i r_i=1$.  The
rescaling defined above will relate two systems with different $r_{\ell}$
in that the densities will be identical up to a rescaling.  However
this rescaling only removes one degree of freedom from the two independent
variables, so unlike $k=2$, one cannot necessarily scale one theory
into another.  Considering $r_0$ and $r_1$, we find
\eqn\rzero{r_0(b)=(1-b)^2+b(2b-1)\bar r_0+b(1-b)\bar r_1}
\eqn\rone{r_1(b)=(3-b)(b-1)+2b(1-b)\bar r_0+b^2\bar r_1}
where $\bar r_0,\bar r_1$ are the values of $r_0,r_1$
prior to rescaling.  Consider the system which is
purely $\ell=0$, or $\bar r_0=1$, $\bar r_1=\bar r_2=0$.  For any $b\neq 1$
then $r_1(b)<0$, which implies that there is no combination of systems
with different $\ell$ which is equivalent to $\ell=0$ up to a rescaling.
This is not the case for the pure $\ell=1$ system.  This system can be
rescaled from $b=1$ to $b=3/4$.  At the latter point one has
$r_0=1/4$, $r_1=0$, and $r_2=3/4$, so this combination of systems,
with an initial
density of $3n_0/4$, will give exactly $3/4$ the density of the $\ell=1$
system at all times.  Similarly, starting with $\bar r_2=1$ the system
can be rescaled from $b=1$ to $b=3/2$.  At the latter point $r_0=1/4$,
$r_1=3/4$, and $r_2=0$.

It should be noted that the correlation function will not be identical
up to a rescaling for any of the systems described above.  This is a
consequence of the fact that the correlation function contains both $\psi$
and $\psi^2$ pieces.

While the reaction considered here is not as generally interesting
as that of $A+B\goto\emptyset$,
it is a suitable starting point for developing the application of RG
methods to these systems.  A similar approach may be applicable to
the reaction
$mA+nB\goto\emptyset$, a system where the universality classes appear
to depend on the nature of the initial conditions \hbox{\refs{\Schnorer
\CDCi\CDCii-\CD}}.

\bigbreak\bigskip\bigskip{\centerline{\bf Acknowledgements}}\nobreak
The author would like to thank J. Cardy for suggesting this problem and
for many helpful conversations.
This work was supported by NSF grant no.~PHY 91-16964 and by
a grant from the SERC.

\appendix{A}{Symmetry Factors}

Diagrams which contain the classical density or the classical response
function are representations of infinite sums of diagrams.  While they
resemble ordinary perturbation theory diagrams, they differ in
combinatorics.  When calculating the Wick contraction
combinatorics one treats propagators as distinguishable,
although the resulting combinatoric factor is then cancelled by a factor
which is absorbed into the definition of the coupling constant.
Our diagrams differ from this in two ways.  First, the classical
density is attached to vertices as an indistinguishable object.  This
will be demonstrated below.  Second, we have chosen to introduce in
the coupling constants no pre-adjusted combinatoric factor.  This is
merely a matter of convention, and is
motivated by the indistinguishability mentioned above, and by the
direct relation of the coupling constant to the parameters used in the
master equation.

The indistinguishability of the density lines can be demonstrated by
considering the contraction of $k$ $\psi$'s, representing a vertex, with
the infinite sum which is the initial state.
\eqn\sfi{\eqalign{\dl\psi^k\dr_{cl}
&=\sum_{m=0}^{\infty}{n_0^m\over m!}\langle\psi^k\bpsi^m\rangle_{cl}\cr
&=\sum_{m=0}^{\infty}{1\over m!}\sum_{\ss m_1,\dots,m_k\atop\ss
m_1+\dots+m_k=m}C^m_{m_1,\dots,m_k}
\prod_{i=1}^k\biggl(n_0^{m_i}\langle\psi\bpsi^{m_i}\rangle_{cl}\biggr).\cr}}
where $C^m_{m_1,\dots,m_k}=m!/(m_1!\dots m_k!)$ is the number of
ways to partition $m$ objects into $k$ distinct boxes.  The sums can
be replaced with unrestricted sums over $m_1\dots m_k$, and the above
expression factors completely, giving
\eqn\sfii{\dl\psi^k\dr_{cl}=\dl\psi\dr_{cl}^k}
The significance of \sfii\ is that there is no $k!$ prefactor.  The $k$
classical density lines which are connected to the vertex are effectively
indistinguishable.

In calculating the classical response function it is necessary to consider
attaching one propagator and $k-1$ density lines to a $\psi^k$ vertex.  This
brings in a factor of $k$, for the number of distinguishable ways the
propagator can be attached.  The remaining $k-1$ densities follow through the
same combinatorics as that shown above, and contribute a factor of $1$.

In general, where the classical response function appears in a diagram it
can be treated as a propagator for combinatorics.  The exception to this
situation is in diagrams such as \two\ and \three, diagrams (d).  Here
the symmetry of the two disconnected branches will result in the branches
attaching as indistinguishable objects.

\listrefs

\listfigs

\bye